\documentclass{emulateapj}
\usepackage{float}
\usepackage{graphicx}

\begin{document}

\title{The Observed Relation between Stellar Mass, Dust Extinction and Star Formation Rate in Local Galaxies}

\author{H.J. Zahid$^{1,2}$, R.M. Yates$^{3}$, L.J. Kewley$^{1,4}$ \& R.P. Kudritzki$^{1,3}$}
\affil{$^{1}$University of Hawaii at Manoa, Institute for Astronomy - 2680 Woodlawn Dr., Honolulu,  HI 96822, USA}
\affil{$^{2}$SAO Predoctoral Fellow, Harvard-Smithsonian Center for Astrophysics, Cambridge, MA 02138, USA}
\affil{$^{3}$Max-Planck-Institute for Astrophysics - Karl-Schwarzschild-Str. 1, D-85741 Garching, Germany}
\affil{$^{4}$Australian National University, Research School of Astronomy and Astrophysics - Cotter Road, Weston Creek, ACT 2611, Australia}

\begin{abstract}

In this study we investigate the relation between stellar mass, dust extinction and star formation rate (SFR) using $\sim \!150,000$ star-forming galaxies from the SDSS DR7. We show that the relation between dust extinction and SFR changes with stellar mass. For galaxies at the same stellar mass dust extinction is \emph{anti-}correlated with the SFR at stellar masses $<10^{10} M_\odot$. There is a sharp transition in the relation at a stellar mass of $10^{10} M_\odot$. At larger stellar masses dust extinction is \emph{positively} correlated with the SFR for galaxies at the same stellar mass. The observed relation between stellar mass, dust extinction and SFR presented in this study helps to confirm similar trends observed in the relation between stellar mass, metallicity and SFR. The relation reported in this study provides important new constraints on the physical processes governing the chemical evolution of galaxies. The correlation between SFR and dust extinction for galaxies with stellar masses $>10^{10} M_\odot$ is shown to extend to the population of quiescent galaxies suggesting that the physical processes responsible for the observed relation between stellar mass, dust extinction and SFR may be related to the processes leading to the shut down of star formation in galaxies.

\end{abstract}


\section{Introduction}

Dust is a fundamental constituent of galaxies. Dust is formed from processed stellar material returned back to the interstellar medium (ISM) either through supernovae or stellar winds. Massive stars ($\gtrsim 8 M_\odot$) which end their lives as Type II supernovae (SNe) and the AGB phase of intermediate mass stars ($1\lesssim M_\odot \lesssim 8$) are considered to dominate stellar dust production in star-forming galaxies while dust in the ISM may also be formed \emph{in situ} from accretion of enriched gas processed by stars \citep{Dwek1998}. Dust is formed from metals and therefore, not surprisingly, a strong correlation is observed between dust and the gas-phase oxygen abundance both in the local universe \citep{Heckman1998, Boissier2004, Asari2007, Garn2010b, Xiao2012, Zahid2012b} and at high redshifts \citep{Reddy2010}. 

\citet{Lequeux1979} first observed a relation between stellar mass and gas-phase oxygen abundance in star-forming galaxies. Using $\sim53,000$ galaxies, \citet{Tremonti2004} have since established a tight relation between stellar mass and the gas-phase oxygen abundance for star-forming galaxies in the local universe. This so-called mass-metallicity (MZ) relation has been observed at low stellar masses \citep{Lee2006, Zahid2012a} and out to high redshifts \citep{Savaglio2005, Erb2006b, Cowie2008, Maiolino2008, Mannucci2009, Lamareille2009, Zahid2011a, Moustakas2011, Yabe2012}. The metallicity is strongly correlated with stellar mass and the shape of the MZ relation is relatively constant with redshift. Over cosmic time the metallicities of galaxies at a fixed stellar mass evolve as galaxies become more enriched at late times. 

The MZ relation is shaped by several important physical processes. Oxygen, the most abundant metal in the ISM, is primarily produced in massive stars which end their lives as Type II SNe, subsequently returning enriched material back to the ISM. However, the observed gas-phase oxygen abundance is also subject to large scale gas flows. Pristine inflowing gas and enriched outflows can both reduce the gas-phase abundance within a galaxy. In pristine inflows the metal content is diluted whereas outflows physically remove metals from the ISM. If the outflowing gas is enriched to levels beyond the ambient ISM, either due to the direct escape of metal-rich ejecta from SNe or to preferential entrainment of metals in galactic winds, the average galaxy metallicity will decline. \citet{Tremonti2004} argue that the enriched outflows which more easily escape the shallow potential wells of low mass galaxies are responsible for the observed MZ relation. Because both inflows and outflows have a similar observational consequence, it has proven difficult to disentangle the effects of gas flows from observations of metallicity alone \citep[see][]{Dalcanton2007}. In this context, the dust content of galaxies may provide important leverage in breaking the degeneracy of these two effects owing to the fact that the observed extinction in galaxies is dependent only on the amount of dust along the line-of-sight and cannot be diluted by inflows of pristine gas. In this study we examine the relation between stellar mass, dust extinction and SFR for star-forming galaxies in the local universe in order to better constrain the physical processes responsible for chemical evolution of galaxies.

In addition to the MZ relation, a tight relation between the stellar mass and SFR of galaxies is observed to exist out to $z\sim2.5$ \citep{Noeske2007a, Elbaz2007, Daddi2007, Pannella2009, Whitaker2012}. The slope and scatter of the stellar mass-SFR (MS) relation are constant and independent of redshift and the overall normalization evolves such that at a fixed stellar mass galaxies at later times have lower SFRs. The fixed slope and scatter in the MS relation suggest that quiescent processes such as cosmological gas accretion are largely responsible for stellar mass growth since $z\sim2.5$. Understanding how the scatter in the MS relation is populated and how quiescent galaxies move off the MS relation will provide important constraints for galaxy evolution. In this study we examine the dust properties of galaxies along the MS relation to shed light on this issue.

The MZ relation and its second parameter dependencies have been investigated by several groups. Most notable is the relation between stellar mass, metallicity and SFR. \citet{Ellison2008} show that there exists a correlation between metallicity and specific star formation rate (sSFR) for galaxies at a fixed stellar mass. The relationship between stellar mass, metallicity and SFR was subsequently investigated by \citet{Mannucci2010} who show that at a fixed stellar mass the SFR is \emph{anti}-correlated with metallicity. They argue for a ``fundamental metallicity relation" between stellar mass, metallicity and SFR. The lower metallicities observed in star-forming galaxies at early times are balanced by the higher SFRs in these galaxies such that the ``fundamental metallicity relation" does not evolve out to $z\sim2$. \citet{Lara-Lopez2010} have independently found a ``fundamental plane" relating the stellar mass, metallicity and SFR of galaxies which appears to match the observational data to $z\sim3.5$.

The observed relation between stellar mass, metallicity and SFR does depend on methodology and sample selection. \citet{Yates2012} reexamine the ``fundamental metallicity relation" and find that while at lower stellar masses the SFR is \emph{anti-}correlated to metallicity, the relation reverses at higher stellar masses such that a \emph{positive} correlation is observed. \citet{Yates2012} argue that the ``twist" in the relation is a result of gas-rich mergers at higher stellar masses which fuel a starburst leading to gas exhaustion and quenching of star formation. Subsequent gas accretion at levels too low to efficiently form large amounts of stars leads to metallicity dilution in these systems thus giving rise to the observed correlation.

In order to shed light on the stellar mass, metallicity and SFR relation and to understand the physical properties of galaxies populating the MS relation we examine the relation between stellar mass, dust extinction and SFR. In Section 2 we describe our sample and in Section 3 we present our results. We provide a detailed discussion of selection and aperture effects in Section 4. In Section 5 we provide a brief discussion and in Section 6 we summarize the main results of the paper. Throughout this work we adopt the standard cosmology $(H_{0}, \Omega_{m}, \Omega_{\Lambda}) = (70$ km s$^{-1}$ Mpc$^{-1}$, 0.3, 0.7) and a \citet{Chabrier2003} IMF.


\section{Data and Methods}

\begin{figure*}
\begin{center}
\includegraphics[width=2\columnwidth]{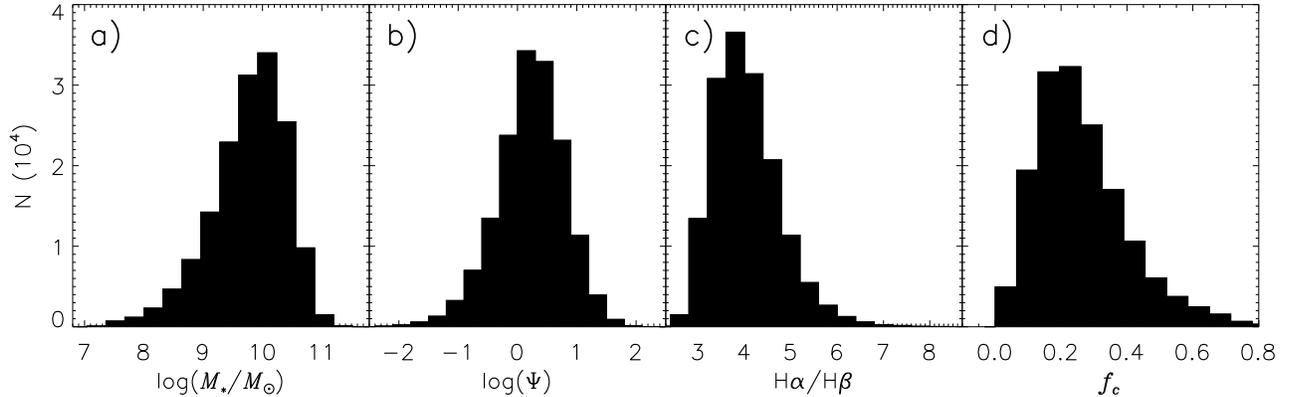}
\end{center}
\caption{The distribution of a) stellar mass, b) SFR ($M_\odot$ yr$^{-1}$), c) Balmer decrement and d) $g$-band fiber covering fraction for the SN8 sample.}
\label{fig:hist}
\end{figure*}


We draw our sample from the SDSS DR7 which consists of $\sim870,000$ unique galaxies spanning a redshift range of $0 < z < 0.7$ \citep{Abazajian2009}. The survey has a Petrosian limiting magnitude of $r_P$ = 17.8 covering 8,200 $deg^2$. The spectra have a nominal spectral range of 3900 - 9100$\mathrm{\AA}$ and a spectral resolution of R $\sim$ 2000. Both the stellar masses, which are determined from the $ugriz$-band photometry \citep{Stoughton2002}, and the emission line fluxes are measured by the MPA-JHU group\footnote{http://www.mpa-garching.mpg.de/SDSS/DR7/}. We adopt the DR7 values in this work but subtract 0.2 dex from the stellar masses for consistency with our previous work where stellar masses are estimated using a different set of routines \citep[see][]{Zahid2011a}. The SFRs in the DR7 are derived using the technique of \citet{Brinchmann2004} with additional improvements given by \citet{Salim2007}. The SFRs are determined from fitting prominent emission lines in the spectra with the largest contribution coming from H$\alpha$ and H$\beta$ and are corrected for dust and aperture effects. 

The emission line fluxes are measured by the MPA/JHU group \citep[see][]{Tremonti2004}. Balmer absorption is prominent in the atmospheres of A stars. For integrated spectra of galaxies a correction for Balmer absorption is required in order to estimate the emission line strength of Balmer lines. The emission lines are continuum subtracted and corrected for stellar absorption by fitting a linear combination of the Charlot \& Bruzual 2008 stellar population synthesis models (Charlot \& Bruzual, in prep). We have scaled the emission line uncertainties of H$\alpha$ and H$\beta$ by 2.473  and 1.882, respectively, as recommended by the MPA/JHU group.

From the parent sample, we select a pure star-forming sample of emission line galaxies. We first distinguish star-forming galaxies from AGN by constraining the ionizing radiation source using the [OIII]$\lambda5007$, [NII]$\lambda6584$, H$\beta$ and H$\alpha$ emission lines \citep{Baldwin1981, Kauffmann2003, Kewley2006}. Following \citet{Kewley2006}, we remove galaxies where 
\begin{equation}
\mathrm{log([OIII]/H\beta)} > 0.61/\left(\mathrm{log([NII]/H\alpha)} - 0.05 \right) + 1.3.
\end{equation}
This selection yields a sample of $388,000$ galaxies. In order to obtain a robust estimate of the Balmer decrement we require that the signal-to-noise (S/N) of the H$\alpha$ and H$\beta$ line be greater than 8. These selection criteria yield a sample of $\sim157,000$ star-forming galaxies. We refer to this selected sample as the SN8 sample.

\citet{Groves2012} find that the H$\beta$ equivalent widths and line fluxes may be systematically underestimated due to an overcorrection for H$\beta$ absorption. They argue that this systematic underestimate in the H$\beta$ line flux leads to a 0.1 mag overestimate of $A_v$. In this study we are investigating the relation between stellar mass, dust extinction and SFR. \citet{Groves2012} conclude that the H$\alpha$ and H$\gamma$ lines do not suffer from the same systematic errors in the absorption correction. In order to asses whether the effect of this possible systematic uncertainty qualitatively changes the relation between stellar mass, dust extinction and SFR, we also determine dust extinction from the H$\alpha$/H$\gamma$ ratio. We require a S/N$ > 3$ in the H$\gamma$ line when determining the dust extinction from the H$\alpha$/H$\gamma$ ratio. Most galaxies in the SN8 sample ($>99\%$) have a S/N $>3$ in H$\gamma$. We determine the dust extinction from the H$\alpha$/H$\gamma$ ratio in Section 4.1 in order to assess any systematic effects due to improper subtraction of H$\beta$ Balmer absorption.

We measure dust extinction from the Balmer decrement. For case B recombination with electron temperature T$_e$  = 10$^4$K and electron density $n_e = 10^2$ cm$^{-3}$, the intrinsic H$\alpha$/H$\beta$ and H$\alpha$/H$\gamma$ ratio are expected to be 2.86 and 6.11, respectively \citep{Hummer1987}. We obtain the intrinsic color excess, E(B$-$V), and the correction for dust attenuation using the extinction law of \citet{Cardelli1989} and a corresponding $R_v = 3.1$. We note that the results of this study are largely independent of our choice of extinction law as the relation presented only relies on relative values of extinction. From the color excess we determine the visual extinction measured in magnitudes from $A_v = R_v$ E(B$-$V). 

In Figure \ref{fig:hist} we show the distribution of stellar masses, SFRs, Balmer decrement and fiber covering fractions for the SN8 sample. We estimate the $g$-band fiber covering fraction, $f_c$, by comparing the photometric and fiber $g$-band magnitude. The $g$-band covering fraction is an estimate of the fraction of the galaxy luminosity contained within the fiber. The median covering fraction for the SN8 sample is 0.24 (see Figure \ref{fig:hist}d). We determine that selection and aperture effects do not significantly bias the observed relation between stellar mass, dust extinction and SFR presented below. In Section 4 we discuss selection and aperture effects in detail.

\section{The Dust and Metallicity Properties of Star-Forming Galaxies}

In Section 3.1 we present the relation between stellar mass, dust extinction and SFR. For comparison, we present the relation between stellar mass, metallicity and SFR in Section 3.2.

\subsection{The Stellar Mass, Dust Extinction and SFR Relation}

\begin{figure*}
\begin{center}
\includegraphics[width=2\columnwidth]{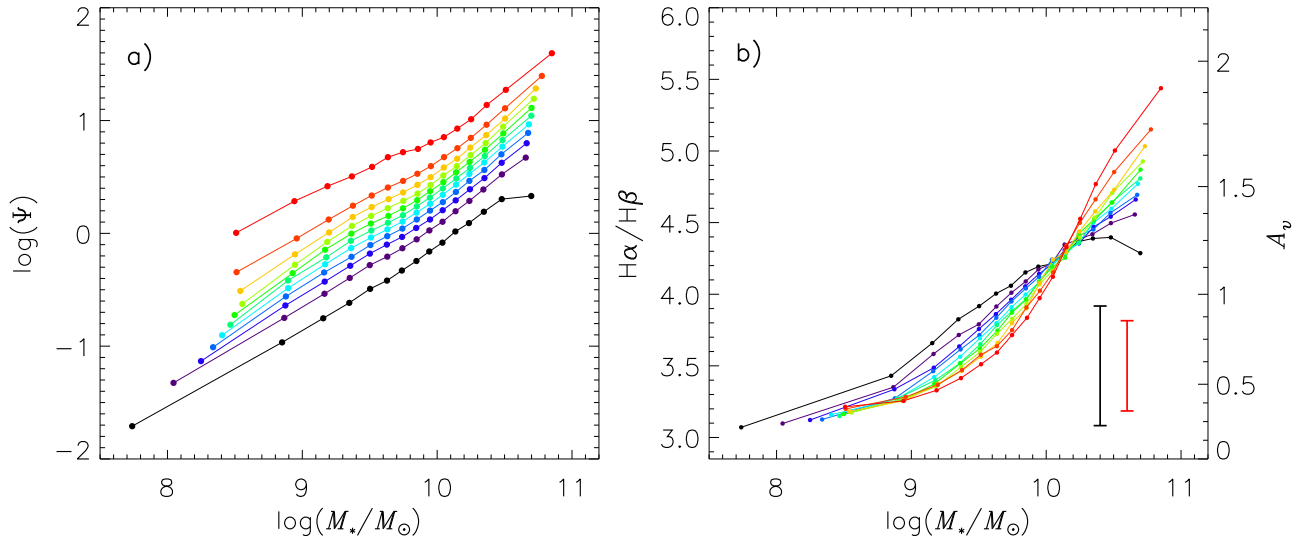}
\end{center}
\caption{The observed relation between stellar mass, dust extinction and SFR ($M_\odot$ yr$^{-1}$). a) Undeciles of the SFR as a function of stellar mass. b) The median Balmer decrement and visual extinction (in magnitudes, see text for details) sorted into bins of stellar mass and SFR. The colors correspond to undeciles of the SFR shown in a). The black error bars show the median 1$\sigma$ dispersion of the data in each bin and the red error bars show the observational uncertainty.}
\label{fig:tau}
\end{figure*}

In Figure \ref{fig:tau} we show the relation between stellar mass, dust extinction and SFR. Hereafter, we refer to the stellar mass, dust extinction and SFR relation as the MDSR. The data are first sorted into 16 equally populated bins of stellar mass and then each mass bin is sorted into 11 equally populated bins of SFR. Each bin contains $\sim890$ galaxies. In Figure \ref{fig:tau}a the different color curves correspond to undeciles\footnote{Each of eleven equal groups into which a population can be divided according to the distribution of values of a particular variable.} of the SFR as a function of stellar mass. In Figure \ref{fig:tau}b we show the median dust extinction sorted into bins of stellar mass and SFR. Again the curves are color coded to match the undeciles of SFR shown in Figure \ref{fig:tau}a (e.g. red curves correspond to the median dust extinction in the highest SFR bin and the black curve the median dust extinction in the lowest SFR bin in each stellar mass bin, respectively). The median 1$\sigma$ scatter of the Balmer decrement within each bin is 0.42 with 0.32 attributable to observational uncertainty. Given the large number of data points within each bin, the median standard error of the Balmer decrement for each bin is 0.01.

There are several notable features present in Figure \ref{fig:tau}. There is a general trend for the extinction to increase with stellar mass \citep[e.g.][]{Brinchmann2004}. Older, higher stellar mass galaxies typically have greater extinction which is most likely due to the greater number of stars in these galaxies evolving through the AGB phase or ending their lives as supernovae \citep[see][]{Dwek1998}. Perhaps more interesting is the relation between extinction and SFR at a fixed stellar mass. At stellar masses $<10^{10}M_\odot$ the dust content of galaxies is \emph{anti}-correlated with the SFR such that galaxies with high SFRs tend to have less dust extinction. At a stellar mass of $\sim \!10^{10} M_\odot$ (or $A_v \sim1.2$) there is a sharp transition and at larger stellar masses the extinction is \emph{positively} correlated with the SFR. Figure \ref{fig:tau} shows two projections of the 3-dimensional MDSR. The reversal of the trend between SFR and dust extinction at a fixed stellar mass can be thought of as a twist in the two dimensional surface defining the relation between stellar mass, dust extinction and SFR.

\begin{figure*}
\begin{center}
\includegraphics[width=2\columnwidth]{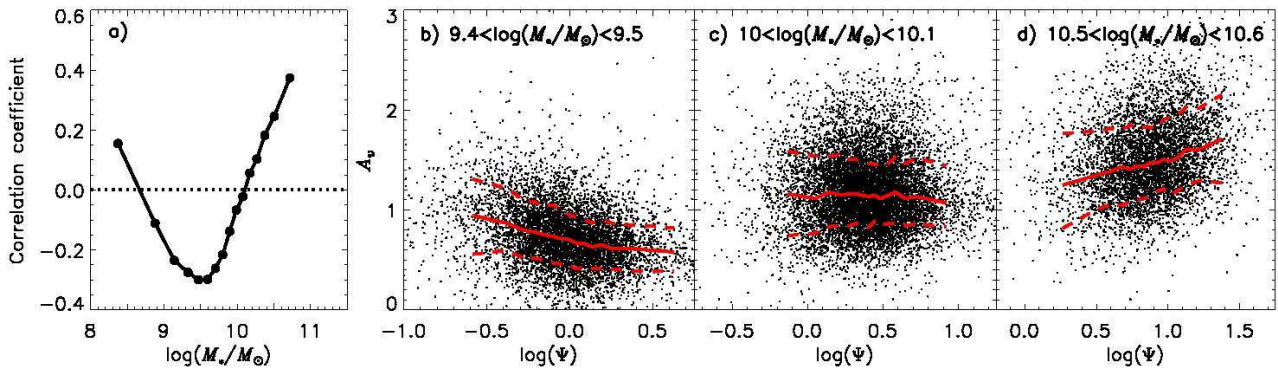}
\end{center}
\caption{a) The Spearman rank correlation coefficient between SFR ($M_\odot$ yr$^{-1}$) and dust extinction in 16 equally populated bins of stellar mass. The dotted line marks the zero point. b-d) Dust extinction plotted as a function of SFR for galaxies in three of the stellar mass bins shown in a). The range of stellar masses are shown in the text of each panel. The red curves are the median dust extinction in 15 bins of SFR and the dashed curves are the 68\% contours of the data.}
\label{fig:cc}
\end{figure*}

The twist is also present in the data without binning in SFR. In Figure \ref{fig:cc} we plot the correlation coefficient between SFR and dust extinction in bins of stellar mass. The data are binned into 16 equally populated bins of stellar mass, same as in Figure \ref{fig:tau}. In the stellar mass range of $8.5 \gtrsim \mathrm{log}(M_\ast/M_\odot) \gtrsim10$ the data show a negative correlation. At $10^{10} M_\odot$ there is a transition to a positive correlation. We demonstrate this visually for the unbinned data in Figure \ref{fig:cc}b-d by showing the relation between dust extinction and SFR in three of the stellar mass bins. At the lowest stellar masses there is evidence for another transition to a positive correlation but the sparsity of data at low stellar masses does not allow us to draw any strong conclusions.

Both \citet{Garn2010b} and \citet{Xiao2012} have studied the dust properties of star-forming galaxies in the SDSS. Neither of these studies report the observed twist in the relation between stellar mass, dust extinction and SFR. However, these studies do not examine the relation between dust extinction and SFR at a fixed stellar mass. Both studies employ a principal component analysis (PCA) on the full sample. PCA is a useful statistical technique for assessing the relative contribution of various correlated parameters. PCA finds the orthogonal linear combinations of the variables that maximize the variance in the data. However, traditional PCA techniques assume a linear relation between the variables. The twist is not obvious using PCA because the relation of dust extinction and SFR is opposite at lower stellar masses as compared to higher stellar masses, thus ``canceling out" in PCA. PCA performed on a sub-sample of the data in restricted range of stellar mass does reveal the twist between dust extinction and SFR (see Figure \ref{fig:cc}).

\subsection{The Stellar Mass, Metallicity and SFR Relation}

\begin{figure}
\includegraphics[width=\columnwidth]{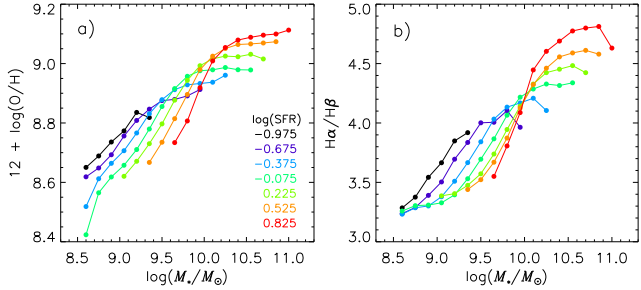}
\caption{The observed relation between a) stellar mass, metallicity and SFR \citep[$M_\odot$ yr$^{-1}$, c.f. Figure 1 of][]{Yates2012} and b) stellar mass, Balmer decrement and SFR using the T2 sample from \citet{Yates2012}. The data are the a) mean metallicities and b) mean Balmer decrements in constant width bins of stellar mass and SFR. The curves are color-coded corresponding to the different SFR bins shown in the legend (the value given for the SFR is the bin center).}
\label{fig:yates}
\end{figure}

In Figure \ref{fig:yates}a we show the relation between stellar mass, metallicity and SFR. Hereafter we refer to this relation as the MZSR. The MZSR is determined using the sample and methodology of \citet{Yates2012}. The data are binned into mass bins of width 0.15 dex and SFR bins of width 0.3 dex. The mean metallicity determined from the Bayesian method of \citet{Tremonti2004} are plotted for each bin and the different color curves correspond to the various SFR bins. The center of each SFR bin is given in the legend of the figure. We refer the reader to \citet{Yates2012} for more details on methodology and sample selection. 

We present Figure \ref{fig:yates} to draw attention to the qualitative similarities in the observed MZSR as compared to the observed MDSR (Figure \ref{fig:tau}). At a fixed stellar mass there exists an \emph{anti-}correlation between the metallicity and SFR for galaxies with stellar masses $\lesssim10^{10}M_\odot$. At a stellar masses $\gtrsim 10^{10} M_\odot$ the trend reverses and a \emph{positive} correlation is observed between metallicity and SFR at a fixed stellar mass. As can be seen in Figure \ref{fig:yates}b, the twist in the MDSR is also present in the \citet{Yates2012} data. 

The \emph{anti}-correlation between metallicity and SFR at lower stellar masses is significantly stronger than the \emph{anti}-correlation between dust extinction and SFR. We determine the correlation coefficient between SFR and metallicity for $\sim7400$ galaxies in the stellar mass range of $9.4<\mathrm{log}(M_\ast/M_\odot)<9.5$. The metallicities are taken from the DR7 and are determined using the Bayesian technique of \citet{Tremonti2004}. The sample correlation coefficient between the aperture corrected SFR and metallicity is $r = -0.41$. Using the SFR determined from the observed H$\alpha$ luminosity in the fiber the Spearman rank correlation coefficient between SFR and metallicity is $r = -0.15$. The \emph{anti-}correlation between stellar mass and metallicity is still present when using H$\alpha$ fiber SFRs \citep[e.g.][]{Mannucci2010}, however the strength of the correlation is diminished. 

Metallicity is a measure of oxygen relative to hydrogen whereas dust extinction is dependent on the absolute number of absorbers within the line of sight. To first order, the observed dust extinction, unlike metallicity, is independent of the gas fraction. The stronger correlation between SFR and metallicity as compared to SFR and dust extinction and much of the difference in the MZSR as compared to the MDSR seen in Figure \ref{fig:yates} is likely due to a correlation between the gas fraction and SFR. Higher gas fractions may sustain higher SFRs while also diluting the metallicity, thus strengthening the \emph{anti-}correlation between metallicity and SFR observed at stellar masses $<10^{10} M_\odot$. Measurements of gas masses in a large sample of star-forming galaxies should provide important insight into the relationship between metallicity and dust.

\section{Systematic, Selection and Aperture Effects}

In this section we investigate possible systematic issues with improper subtraction of H$\beta$ absorption (Section 4.1), biases in the observed MDSR associated with our method of sample selection (Section 4.2) and systematic effects of measuring global physical properties of galaxies from emission lines observed within a limited aperture (Section 4.3). We conclude that selection and aperture effects are not significant in our determination of the MDSR. 

\subsection{Systematic Effects in the H$\beta$ Absorption Correction}

\begin{figure}
\begin{center}
\includegraphics[width=\columnwidth]{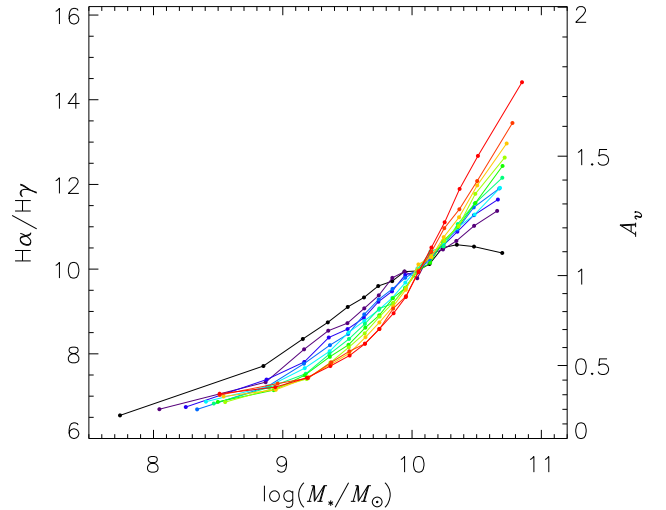}
\end{center}
\caption{The observed relation between stellar mass, dust extinction and SFR ($M_\odot$ yr$^{-1}$). Similar to Figure \ref{fig:tau}b but with dust extinction determined from the H$\alpha$/H$\gamma$ ratio.}
\label{fig:gamma}
\end{figure}

In Figure \ref{fig:gamma} we plot the MDSR with dust extinction determined from the H$\alpha$/H$\gamma$ ratio. The relation presented in Figure \ref{fig:gamma} displays the same characteristics as Figure \ref{fig:tau}b. An \emph{anti-}correlation between dust extinction and SFR at stellar masses $<10^{10}M_\odot$ and a positive correlation at higher stellar masses. We observe a $\sim0.1$ magnitude greater extinction when determining $A_v$ from H$\alpha$/H$\beta$ as compared to H$\alpha$/H$\gamma$. \citet{Groves2012} find a similar offset and by comparing SDSS DR7 data with DR4 data. They attribute the difference in extinction determined from H$\alpha$/H$\beta$ and H$\alpha$/H$\gamma$ to a systematic error in the subtraction of the underlying H$\beta$ Balmer absorption. While systematic effects in subtracting the underlying H$\beta$ absorption may affect the absolute measurement of $A_v$, we conclude that the observed twist in the MDSR is not affected. A comparison of Figure \ref{fig:gamma} with Figure \ref{fig:tau}b shows that a greater difference in $A_v$ is observed at higher stellar masses and SFRs. The overestimation of $A_v$ appears to be correlated with the stellar mass and SFR.

\subsection{Selection Effects}

We select star-forming galaxies from the parent sample using the BPT method which allows us to identify and remove galaxy spectra dominated by AGN emission (see Section 2). We require a S/N $>$ 8 in the H$\alpha$ and H$\beta$ emission line fluxes in order to obtain a robust estimate of the SFR and Balmer decrement from which we measure the extinction. The strength of the Balmer lines scales with number of UV ionizing photons originating from massive stars and therefore is a good indicator of the SFR \citep{Kennicutt1998b}. We may bias our sample by selecting galaxies above a fixed S/N threshold because galaxies with low levels of star formation will have weak Balmer lines that may not meet our S/N requirement. This selection criteria could lead to a spurious MDSR if the bias is mass dependent.

\begin{figure}
\begin{center}
\includegraphics[width=\columnwidth]{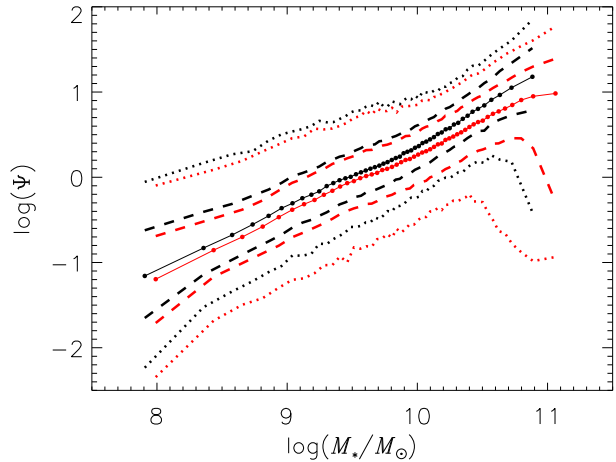}
\end{center}
\caption{The SFR ($M_\odot$ yr$^{-1}$) distribution for the SN8 (black curves) and SN3 (red curves) samples. The median SFR in 30 equally populated bins of stellar mass are shown by the filled circles and solid curves. The 68\% and 95\% contours of the distribution are shown by the dashed and dotted curves, respectively.}
\label{fig:sfr_comp}
\end{figure}

In order to investigate the bias introduced by our S/N requirement we compare the distribution of SFR as a function of stellar mass for the SN8 sample with a sample selected requiring a S/N $>$ 3 in the H$\alpha$ and H$\beta$ emission lines. We refer to this sample as the SN3 samples. The SN3 sample consists of $\sim259,000$ galaxies and contains a factor of $\sim1.6$ more galaxies than the SN8 sample. In Figure \ref{fig:sfr_comp} we plot the distribution of SFRs in 30 bins of stellar mass for the SN8 (black curves) and SN3 (red curves) samples. The median of the SFR distribution (solid curves) of the SN3 sample is typically $\sim$0.1 dex lower than the SN8 sample except at the highest stellar masses where the difference is larger. 

We are interested in determining the MDSR for the star-forming sequence of galaxies as identified by \citet[and many others]{Noeske2007a}. The population of star-forming galaxies out to $z\sim2$ is characterized by a near unity slope and constant scatter in the relation between stellar mass and SFR that is independent of redshift \citep{Noeske2007a, Elbaz2007, Daddi2007, Pannella2009, Whitaker2012}. In the SN8 and SN3 sample, the scatter in the SFR distribution increases at higher stellar masses and a population of massive, low SFR galaxies is present in the distribution shown in Figure \ref{fig:sfr_comp}. By examining the sersic index of galaxies on the stellar mass-SFR diagram, \citet{Wuyts2011} show that this region of the diagram is populated by quiescent galaxies best described by de Vaucouleurs profiles. Decreasing our S/N threshold slightly broadens and shifts the distribution of SFRs at all stellar masses and selects a greater number of massive, quiescent galaxies.

\begin{figure}
\begin{center}
\includegraphics[width=\columnwidth]{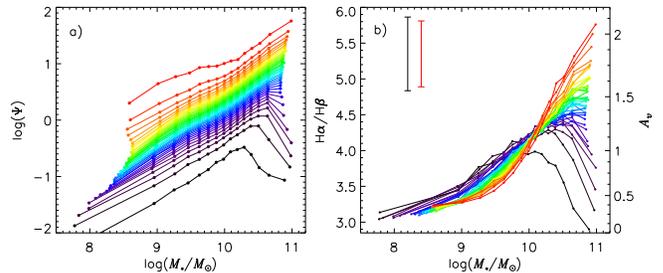}
\end{center}
\caption{The MDSR for the SN3 sample.}
\label{fig:tau_sn}
\end{figure}

In Figure \ref{fig:tau_sn} we reexamine the stellar mass, dust extinction and SFR relation using our SN3 sample. The greatest difference in the MDSR determined from the SN3 sample as compared to the SN8 sample is for galaxies with stellar masses $\gtrsim 10^{10} M_\odot$ and is attributable to the greater number of quiescent galaxies in the sample. We conclude that our S/N selection criterion does not strongly select against the star-forming sequence of galaxies. The S/N criterion of the SN8 sample does not completely remove quiescent galaxies from the sample, though they constitute a small fraction of the sample. Thus, selection bias is not a significant. Comparison with the SN3 sample suggests that the downturn in the relation between stellar mass and dust extinction observed in the highest mass, lowest SFR bin of the SN8 sample is a consequence of the presence of quiescent galaxies in the sample (see Figure \ref{fig:tau}a). This suggests that twist observed in the MDSR and the \emph{anti}-correlation between SFR and dust extinction is related to the shutting down of star formation in galaxies.

We apply a higher S/N threshold in order to emulate selection effects that may be present in high redshift data. The ``twist" in the MDSR is \emph{clearly} observed in the MDSR when the S/N threshold for H$\alpha$ and H$\beta$ emission is $<20$. At a S/N$\gtrsim25$, there is no longer a ``twist" in the MDSR. In the SDSS data 74\% of the galaxies are removed when a S/N$>25$ is required. At a fixed stellar mass this selection criterion preferentially removes low SFR galaxies. This demonstrates that incompleteness in SFR may bias the observed MDSR and care must be taken when investigating the MDSR in higher redshift samples.

\subsection{Aperture Effects}
We investigate aperture effects associated with measurements of the SFR (Section 4.2.1) and the Balmer decrement (Section 4.2.2).

\subsubsection{Star Formation Rate}

The SFRs used in this study are derived by the MPA/JHU group using the technique developed by \citet[hereafter B04]{Brinchmann2004}. B04 model the stellar continuum and absorption lines using \citet{Bruzual2003} stellar population synthesis models. They model the emission lines using the CLOUDY photoionization code \citep{Ferland1996} and \citet{Charlot2001} nebular emission models. Dust attenuation is largely constrained using the Balmer decrement with small contributions from other emission lines. This procedure gives the SFR within the 3'' fiber aperture. B04 apply a fiber aperture correction in order to obtain the total SFR. The correction is derived by calculating the dependency of SFR on color within the fiber. The SFR outside the fiber is accounted for by assuming that the color dependency of the SFR is the same inside and outside the fiber (see B04 for more details).

\citet{Salim2007} derive dust corrected SFRs for $\sim50,000$ galaxies in the local universe by fitting the UV and optical spectral energy distribution (SED) with a library of stellar population synthesis models. The SFRs determined from the SED are not subject to aperture effects. \citet{Salim2007} show that the aperture and dust corrected SFRs derived by B04 for the sample of star-forming galaxies (i.e. those with H$\alpha$ detected with S/N $>$ 3) agree with those determined from the UV and optical SED. There is a systematic difference of 0.02 dex (when 3$\sigma$ outliers are excluded) and the scatter in the two methods is accounted for by the uncertainty of each method. \citet{Salim2007} conclude that for star-forming galaxies the UV-based SFRs agree remarkably well with the aperture corrected SFRs derived by B04. For the star-forming galaxies in the SN8 sample, the SFRs made available in the DR7 and used in this study are robust against aperture bias.

\begin{figure}
\begin{center}
\includegraphics[width=\columnwidth]{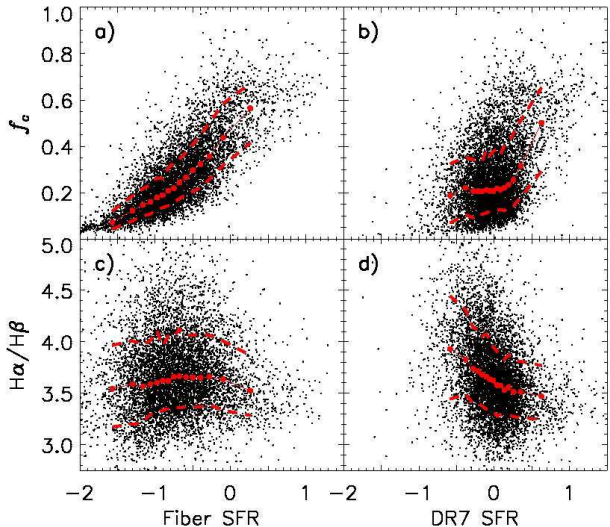}
\end{center}
\caption{Relation between covering fraction, $f_c$, and a) H$\alpha$ SFRs and b) aperture corrected SFRs ($M_\odot$ yr$^{-1}$). The relation between Balmer decrement and c) H$\alpha$ SFR and d) aperture corrected SFRs. The data are the SN8 sample in a limited mass range ($9.4 < \mathrm{log}(M_\ast/M_\odot) < 9.5$). The median covering fraction plotted by the solid red curve in 15 bins of a) H$\alpha$ SFRs and b) aperture corrected SFRs. The median Balmer decrement plotted by the solid red curve in 15 bins of c) H$\alpha$ SFRs and d) aperture corrected SFRs. The 68\% contours of the distributions are shown by the dashed red curves.}
\label{fig:cf}
\end{figure}

The MDSR derived in this study is dependent on how the SFR is measured. To demonstrate the need for an aperture correction, we also determine the SFR from the dust corrected H$\alpha$ luminosity observed in a 3'' fiber aperture using the conversion of \citet{Kennicutt1998b}. We refer to the SFR determined from the observed H$\alpha$ luminosity as the Fiber SFR and explicitly refer to the aperture corrected SFRs as the DR7 SFR. The relation between stellar mass, dust extinction and Fiber SFR is qualitatively different at lower stellar masses when compared to the same relation using DR7 SFRs. At a fixed stellar mass, the Fiber SFRs are not \emph{anti-}correlated with dust extinction at stellar masses $<10^{10}M_\odot$. We attribute the different relation observed between dust extinction and SFR at lower stellar masses to aperture effects resulting from the use of Fiber SFRs rather than the DR7 SFRs. In Figure \ref{fig:cf}a we show the distribution of aperture covering fraction for galaxies as a function of Fiber SFR in a narrow mass range ($9.4<\mathrm{log}(M_\ast/M_\odot)<9.5$). The fiber covering fraction is strongly correlated with the Fiber SFR. In Figure \ref{fig:cf}c we show that Fiber SFRs and Balmer decrements are weakly (positively) correlated at Fiber SFRs $< 0$ but show a negative correlation at higher Fiber SFRs.

In Figure \ref{fig:cf}b and d we show the distribution of covering fraction and Balmer decrement, respectively, plotted as a function of DR7 SFRs. The Balmer decrement is not strongly correlated with DR7 SFRs. This interval contains $\sim$75\% of data. At higher DR7 SFRs there is a positive correlation between the SFR and covering fraction. The correlation between covering fraction and DR7 SFR at higher stellar masses arises because the sample is not volume limited and we have applied a fixed S/N threshold. We observe a similar relation between stellar mass, dust extinction and SFR using the volume limited sample of \citet{Zahid2011a} which is comprised of data selected in a redshift range of $0.04<z<0.1$. No correlation between covering fraction and DR7 SFR is observed in the volume limited sample in the mass range investigated in Figure \ref{fig:cf}. However, restricting the redshift range removes a substantial number of low and high mass galaxies. Because the effect of not selecting a volume limited sample is not significant and does not change our conclusions, we do not apply this additional criterion in selecting data.

By comparing Figure \ref{fig:cf}c and d we show that not including an aperture correction for SFR results in a spurious correlation between Balmer decrement and SFR. Figure \ref{fig:cf} highlights the importance of applying an aperture correction to SDSS SFRs when examining global trends in order to avoid systematic bias in relations between SFR and other physical properties. 

\subsubsection{Balmer Decrement}

Several studies have reported variations in dust extinction with galactocentric radius \citep[e.g.][]{Holwerda2005, Boissier2007, Tamura2009, Munoz-Mateos2009}. However, \citet{Kewley2005} find little evidence for systematic variation between nuclear and global extinction for covering fractions $>20\%$. Here we test possible biases that may result from aperture effects in determining the Balmer decrement.

\begin{figure}
\begin{center}
\includegraphics[width=\columnwidth]{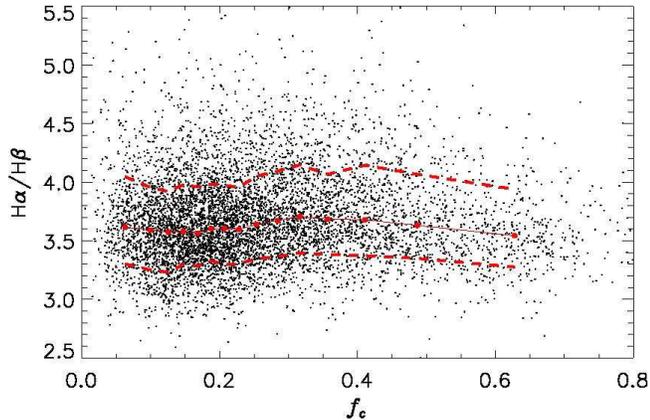}
\end{center}
\caption{The Balmer decrement plotted against the aperture covering fraction, $f_c$. The black data points are taken from the SN8 sample in a limited mass range ($9.4 < \mathrm{log}(M_\ast/M_\odot) < 9.5$). The median Balmer decrement is plotted by the solid red curve in 15 bins of aperture covering fraction, $f_c$. The 68\% contours of the distributions are shown by the dashed red curves.}
\label{fig:dec_cf}
\end{figure}

If strong negative dust gradients exist in galaxies then the measured Balmer decrement will be biased towards larger values of extinction in galaxies with small covering fractions. In this case, the measured Balmer decrement will not reflect the global dust properties. We first test for any bias by plotting the Balmer decrement as a function of the aperture covering fraction. In Figure \ref{fig:dec_cf} we plot the Balmer decrement as a function of aperture covering fraction for data in a narrow mass range. For data in the stellar mass range of $9.4 < \mathrm{log}(M_\ast/M_\odot) < 9.5$, the Balmer decrement is not strongly correlated to the aperture covering fraction.

We also apply a more global test for potential bias in the Balmer decrement due to aperture effects. We divide the SN8 sample into two equally populated subsamples. Sample S1 is comprised of galaxies that have covering fraction less than the median covering fraction of the SN8 sample ($f_c < 0.24$). Sample S2 is the complementary sample with covering fraction greater than the median covering fraction. Each sample has $\sim$74,000 galaxies. 

\begin{figure*}
\begin{center}
\includegraphics[width=2\columnwidth]{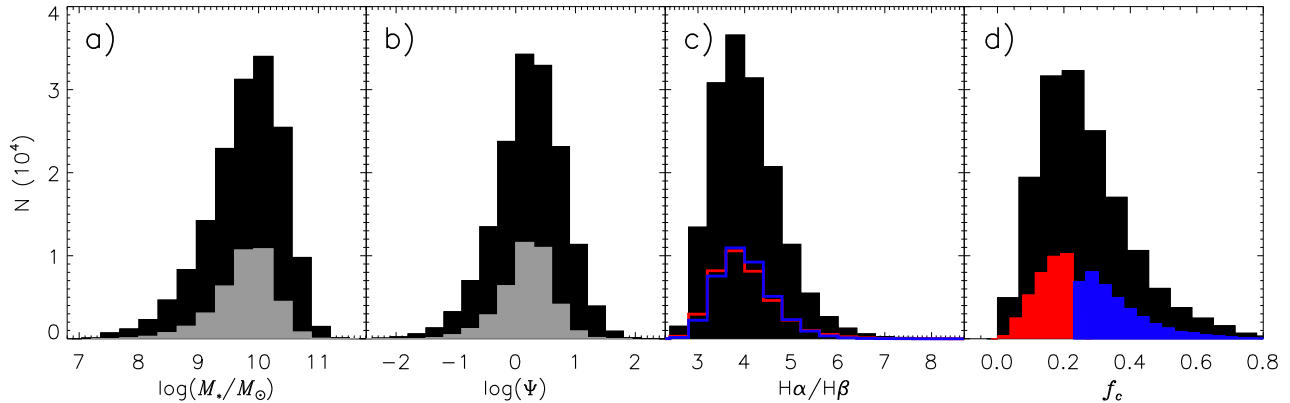}
\end{center}
\caption{The black histograms are for the SN8 sample and are the same as Figure \ref{fig:hist}. The stellar mass and SFR ($M_\odot$ yr$^{-1}$) distribution are identical for the S1s and S2s sample and are shown by the gray histograms. The distribution of the c) Balmer decrement and d) covering for the S1s and S2s samples are shown by the red and blue histograms, respectively.}
\label{fig:dec_dist}
\end{figure*}

We want to examine the bias in the Balmer decrement associated \emph{only} with the covering fraction. While in small bins of stellar mass, the aperture corrected SFR is not strongly correlated to covering fraction (see Figure \ref{fig:cf}b), the SFRs and stellar masses across the whole sample are correlated to the covering fraction. The correlation of the SFR and stellar mass with covering fraction is a consequence of the fact that SDSS is a magnitude limited survey and we have selected our sample using a fixed S/N threshold. Therefore we must take care to remove second order correlations between Balmer decrement and covering fraction resulting from correlations between covering fraction, SFR and stellar mass. We do this by randomly selecting subsets of the S1 and S2 data that are matched to have an identical SFR and stellar mass distributions (in bins of 0.05 dex width). We refer to these as the S1s and S2s subsamples. The S1s and S2s subsamples are comprised of $\sim39,000$ galaxies each. Figure \ref{fig:dec_dist}a and b show the stellar mass and SFR distribution for S1s and S2s in gray. The stellar mass and SFR distributions of the S1s and S2s are representative of the SN8 sample and are identical for the S1s and S2s sample by design. Figure \ref{fig:dec_dist}c and d show the distribution of the Balmer decrement and covering fraction for the S1s (red histogram) and S2s (blue histogram) samples, respectively.

\begin{figure}
\begin{center}
\includegraphics[width=\columnwidth]{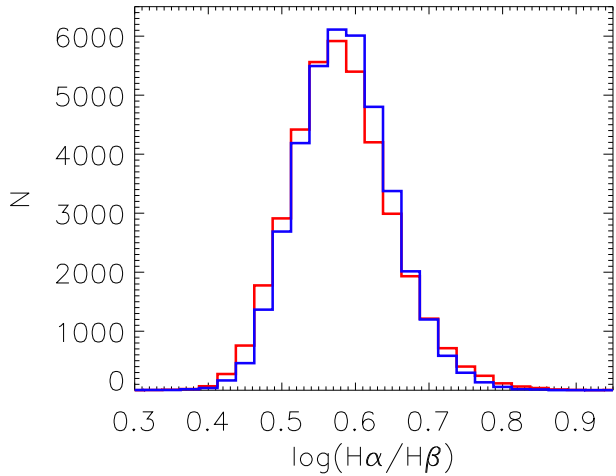}
\end{center}
\caption{Distribution of the logarithm of the Balmer decrement for the S1s (red) and S2s (blue) samples.}
\label{fig:dec_norm}
\end{figure}

\begin{figure*}
\begin{center}
\includegraphics[width=2\columnwidth]{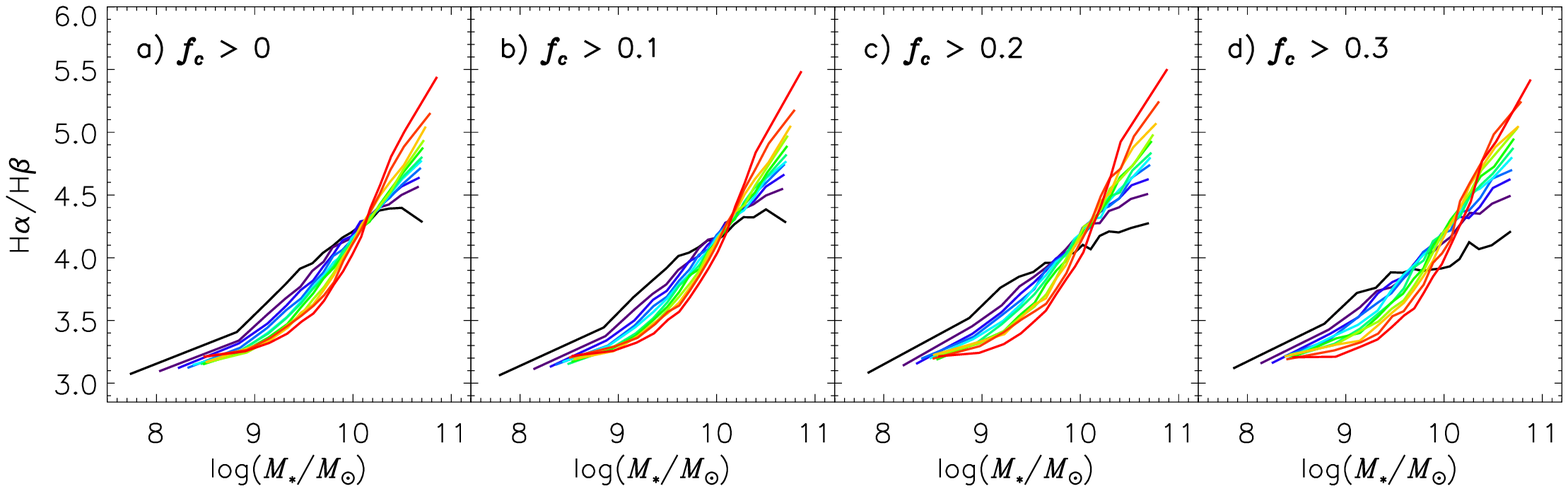}
\end{center}
\caption{The MDSR for the SN8 sample but with a minimum aperture covering fraction requirement of a) 0, b) 0.1, c) 0.2 and d) 0.3.}
\label{fig:cf_relation}
\end{figure*}

In Figure \ref{fig:dec_norm} we show the distribution of the logarithm of the Balmer decrement for the S1s (red histogram) and S2s (blue histogram) samples. The data are nearly log-normally distributed. The median observational uncertainties in the Balmer decrement for the S1s and S2s samples are 0.34 and 0.28, respectively. The S1s sample ($f_c < 0.24$) has a slightly broader distribution as compared to the S2s sample. The greater width of the S1s distribution may be due to the larger observational uncertainties of the S1s sample. Despite the very different distribution in covering fraction, the distribution of Balmer decrement in the S1s and S2s samples are very similar. If there were a strong bias in the measured Balmer decrement due to aperture effects, we would expect a relative shift in the distribution of the Balmer decrement for the S1s and S2s samples. We conclude that aperture effects do not significantly bias the measurement of the dust extinction in the SN8 sample.

The Balmer decrement measured in the fiber is a flux-weighted average over many HII regions. We perform a simple calculation to test whether the similarity in the S1s and S2s Balmer decrement distribution is consistent with the extinction observed in nearby galaxies. \citet{Munoz-Mateos2009} measure the radial attenuation profile for galaxies in the Spitzer Infrared Nearby Galaxies Survey sample. The typical (median) radial attenuation gradient measured is -0.023 mags/kpc. We convert this gradient measured in magnitudes of extinction into a gradient in Balmer decrement assuming the \citep{Cardelli1989} extinction curve. We adopt an exponential profile for H$\alpha$ surface brightness with a fiducial scale length of 4 kpc. Using these values we determine the flux-weighted Balmer decrement as a function of fiber covering fraction. We find that for the fiducial values of radial attenuation gradient and H$\alpha$ scale length, the Blamer decrement is overestimated by $\sim5\%$ when the covering fraction is 5\% (the minimum for the SN8 sample) and a 4\% overestimate when the covering fraction is 20\%. The median covering fraction of the S1s and S2s samples is 16\% and 33\%, respectively. The relative systematic error in the Balmer decrement between these two covering fraction is $\sim1\%$. The calculated values for the relative error do not depend strongly on scale length adopted but do increase if a steeper attenuation gradient is adopted (e.g. relative error of 5\% for an attenuation gradient of -0.1 mags/kpc). We conclude that distribution of Balmer decrement shown in Figure \ref{fig:dec_norm} is consistent with the typical (shallow) gradients found in local spiral galaxies \citep[e.g.][]{Munoz-Mateos2009}.

As we have shown in this section, the measured dust extinction and SFRs are not significantly biased by aperture effects. However, because aperture effects are present in the SDSS sample in a complicated way, we also test whether the observed MDSR is effected by aperture bias. In Figure \ref{fig:cf_relation} we plot the MDSR by applying an increasing minimum aperture covering fraction requirement. In Figure \ref{fig:cf_relation}a-d we apply a minimum covering fraction of 0, 0.1, 0.2 and 0.3 in selecting data and determine the MDSR using $\sim$157,000, $\sim$143,000, $\sim$98,000 and $\sim$52,000 galaxies, respectively. A very similar MDSR is observed in Figure \ref{fig:cf_relation}a-d. We conclude that selection and aperture effects are not significant in our determination of the MDSR.

\section{Discussion}

In this contribution we have presented the MDSR for local star-forming galaxies. The physics of dust production, destruction and evolution are not well understood \citep[for review see][]{Draine2003} and it is beyond the scope of this study to make detailed considerations of these processes. It is well established that dust forms from metals and therefore it is not surprising that a strong correlation between dust and metals is observed \citep[e.g.][]{Garn2010b}. \citet{Zahid2012b} derive a relation between stellar mass, metallicity and dust extinction. They show that the dust extinction increases with \emph{both} stellar mass and metallicity. That is, dust extinction increases with stellar mass and at a fixed stellar mass, the metallicity and dust extinction are positively correlated. The correlation between stellar mass, dust extinction and metallicity is also evident from PCA \citep{Garn2010b, Xiao2012}.

The MDSR shows similar trends as the MZSR. While dust extinction can be easily estimated from the Balmer decrement, metallicity is substantially more difficult physical quantity to measure. Various methods have been developed to determine the gas-phase oxygen abundance from the emission line properties of star-forming galaxies \citep[for review of methods see][]{Kewley2008}. The most commonly used methods rely on theoretically or empirically calibrating ratios of strong collisionally excited emission lines to recombination lines. However, these so-called strong line methods suffer from various systematic uncertainties. In particular, \citet{Kewley2008} show that the metallicity determined for the same sample of galaxies can differ by up to $\sim$0.6 dex depending on the choice of calibration. Thus the observed MDSR presented in this study helps to independently establish the observed MZSR presented in \citet{Yates2012}.

The observed MZ relation and its second parameter dependencies serve as a Rosetta stone for understanding the physical processes responsible for the chemical evolution of galaxies. Because dust is formed from metals, we favor a common physical origin for the twist observed in both the MDSR presented in this study and the MZSR investigated by \citet{Yates2012}. While dust may be destroyed, metals can not and therefore the observed MDSR can not be explained by destruction processes unless the similarities in the MDSR and MZSR are taken to be coincidental. Moreover, given that metals are locked up in dust, we may expect an opposite trend in the MDSR and MZSR if dust destruction is the responsible mechanism since the destruction of dust grains should liberate the constituent metals, thus increasing the gas-phase abundance. The metallicities of galaxies are set by a balance between star formation and gas flows. Dust extinction is, to first order, insensitive to the infall of metal-poor gas and cannot be diluted. Therefore the observed twist in the MDSR and the MZSR can not \emph{solely} be explained as a consequence of gas dilution. We consider differential mass loss in galaxies which is related to the SFR a natural explanation of the observed twist in the MDSR and MZSR. 

\citet{Whitaker2012} study the relation between stellar mass and SFR out to $z=2.5$. They conclude that the dust attenuation observed in star-forming galaxies increases with stellar mass. They find that for galaxies with stellar masses $\gtrsim10^{10} M_\odot$ dusty, blue galaxies populate the upper envelope of the scatter in the MS relation. Conversely, red low-dust galaxies have lower observed SFRs and are interpreted to possibly be in the process of shutting down star formation. They show that the MS relation at $1 < z < 1.5$ for red galaxies has a shallower slope than the MS relation for blue galaxies. It is possible that at lower stellar masses the two relations cross such that a similar trend of a reverse in the correlation between SFR and dust attenuation at a fixed stellar mass may be present in the high redshift population of star-forming galaxies \citep[see Figure 4 of][]{Whitaker2012}. However, due to incompleteness no firm conclusions can be drawn regarding the correlation between SFR and dust attenuation at low stellar masses. 

The results of our study are consistent with the interpretation of \citet{Whitaker2012}. In particular, decreasing our S/N threshold to S/N $>3$ (Section 4.2) results in the inclusion of a significant fraction of quiescent galaxies as identified by the sersic index analysis of \citet{Wuyts2011}. Figure \ref{fig:tau_sn} shows continuity in the correlation between dust extinction and SFR observed at stellar masses $>10^{10} M_\odot$. This suggests that the physical mechanism responsible for the correlation between SFR and dust extinction at higher stellar masses and the twist in the MDSR and MZSR may be related to the physical processes leading to the shut down of star formation and the migration of galaxies to the red sequence. 

\section{Summary}

In this study we have investigated the relation between stellar mass, SFR and dust extinction. We conclude that:
\begin{itemize}
\item{Our analysis is consistent with the conclusions of \citet{Garn2010b} that the strongest correlation in the data is the \emph{positive} correlation between stellar mass, SFR and dust extinction.}

\item{The relation between SFR and dust extinction \emph{at a fixed stellar mass} is mass dependent. At a fixed stellar mass, an \emph{anti}-correlation between the SFR and dust extinction is observed for galaxies with stellar masses $<10^{10}M_\odot$. There is a sharp transition at  a stellar mass of $10^{10}M_\odot$. In galaxies with larger stellar masses there is a \emph{positive} correlation between the SFR and dust extinction at a fixed stellar mass.}

\item{The relation between stellar mass, metallicity and SFR \citep[see][]{Yates2012} shows the same trends as the relation between stellar mass, dust extinction and SFR. Unlike metals, dust can not be diluted by inflows of pristine gas. The observed stellar mass, dust extinction and SFR relation provides important new constraints for understanding the physical processes governing chemical evolution of galaxies.}

\item{Quiescent galaxies are observed to populate the high mass, low SFR part of the stellar mass - SFR diagram. When including quiescent galaxies in the sample we find continuity in the correlation between dust extinction and SFR at stellar masses $>10^{10} M_\odot$. The physical processes responsible for the relation between stellar mass, dust extinction and SFR at the high mass end may be related to the physical processes leading to the shutdown of star formation.}

\end{itemize}

In a forthcoming paper we develop a model incorporating momentum driven outflows as a possible explanation for the observed relation between stellar mass, dust extinction and SFR presented here.

\acknowledgements

We thank the anonymous referee for careful reading of the manuscript and G. Cresci for useful comments. HJZ and LJK gratefully acknowledge support by NSF EARLY CAREER AWARD AST07-48559. RMY acknowledges the financial support of the Deutsche Forschungsgesellshaft (DFG). RPK acknowledges support by the Alexander-von-Humboldt Foundation and the hospitality of the Max-Planck-Institute for Astrophysics in Garching where part of this work was carried out. 
 


 \end{document}